\def\tPm{\tilde{\bm P}_m}

\def\Pm{{\bm P}_m}

\def\FT{{\cal F}}

\def\eps{\varepsilon}
\def\half{\tfrac{1}{2}}
\def\trho{\tilde{\rho}}

\def\figw{0.14\textwidth}
\def\figwIV{0.22\textwidth}

\documentclass[twocolumn,osajnl2,showpacs,preprintnumbers,floats]{revtex4}

\usepackage{amsmath, amsthm, amssymb}

\usepackage {graphicx}
\usepackage {bm}
\usepackage {subfigure}
\newtheorem{thm}{Theorem}[section]

\def\@dotsep{4.5}
\makeatother

\begin{document}

\title{Ab initio compressive phase retrieval}

\begin{abstract}
Any object on earth has two fundamental properties: it is finite, and
it is made of atoms.  Structural information about an object can be
obtained from diffraction amplitude measurements that account for
either one of these traits.  Nyquist-sampling of the Fourier
amplitudes is sufficient to image single particles of finite
size at any resolution. 
Atomic resolution data is routinely used to image molecules
replicated in a crystal structure. Here we report an algorithm that
requires neither information, but uses the fact that an image of a
natural object is compressible. 
Intended applications include tomographic diffractive imaging, 
crystallography, powder diffraction, small angle x-ray scattering 
and random Fourier amplitude measurements.

\end{abstract}

\author{Stefano Marchesini}
\affiliation{
Centre for Free-Electron Laser Science, DESY, Notkestrasse 85, 22607
Hamburg, Germany.
}
\affiliation{Permanent address: Advanced Light Source, Lawrence 
Berkeley National Laboratory, 1 Cyclotron Rd, Berkeley CA 94720, USA. 
e-mail: smarchesini@lbl.gov}



\preprint{XXXXXXXXXX}

\maketitle

\section{Intro}
 In a standard imaging system, light scattered from an object
forms a diffraction pattern which encodes information about the object
Fourier components. A lens recombines the scattered rays so that they
interfere correctly to form an image: it performs an inverse Fourier
transform of the diffraction pattern to convert Fourier (reciprocal)
representation of the object into real space information. 

At visible wavelengths, aberration-free lenses can provide diffraction
limited images within a limited depth of field. Smaller wavelengths
offer much higher resolutions by reducing the diffraction limit.
X-rays also offer the ability to penetrate through thick objects and allow
one to examine elemental, chemical, or magnetic information by exploiting
mechanisms such as resonant X-ray scattering. Unfortunately,
diffraction limited optics are harder to come by for X-rays, whose
paths are difficult to manipulate. For higher resolutions, the optics
need to cope with light scattered to high angles. Building such optics
is a difficult technical challenge. Currently, focal widths of tens of
nanometers are achievable, but lenses capable of atomic resolution are
so far well beyond reach.

Diffraction and scattering experiments overcome this problem by
eliminating the need for any optics. The concept is to record the
scattering pattern created by an object and perform the
re-interference normally done by a lens numerically instead.  Since no
optical elements are used, aberration free images may be obtained with
resolutions limited in principle only by the maximum momentum transfer
which can be achieved.  However, the resulting image is additionally
limited by the computer's ability to recover the entire image from
incomplete Fourier information and loss of phase information.  The
intimate relationship between the phase-front and the direction of
propagation seems to suggest that the task of recombining x-rays back at the sample
position would seem hopeless. It is not so under a surprisingly small
set of conditions.

\begin{figure}
	\includegraphics[width=0.2\textwidth]{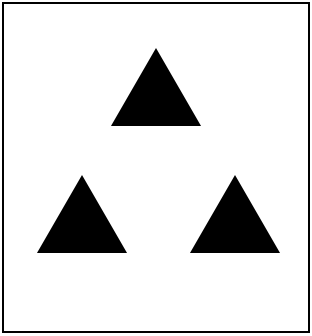}
	\includegraphics[width=0.2\textwidth]{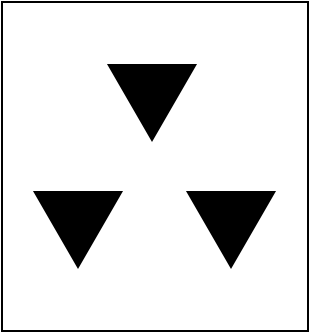}
\caption{Simple homometric objects\cite{homometric}, with the same Fourier
magnitude, are composed by the convolution between two
non-centrosymmetric objects.
\label{fig1}}
\end{figure}

The importance of fine sampling of the diffraction pattern intensity
was recognized at an early stage in x-ray crystallography.  The
observation that Bragg diffraction undersamples the
diffracted intensity pattern \cite{Perutz:1938,Sayre:1952} was 
followed by the demonstration
that the solutions to
Nyquist sampled diffraction patterns are almost always unique 
\cite{bruck, bates, hayes1, hayes2} 
(although one could easily make up examples where this is not the
case -see Fig. \ref{fig1}).

These ideas, along with the development of powerful light sources,
producing collimated beams of coherent x-rays, enabled the development of
coherent x-ray diffraction microscopy \cite{miao:nature} (aka lensless or diffractive
imaging).  This technique aims at imaging, through coherent
illumination, Fourier amplitude measurements and adequate sampling,
macroscopic objects such as entire cellular organisms \cite{shapiro:PNAS05},
 or nanoporous aerogel structures
\cite{barty:PRL08}. See \cite{spence:book} for a review. 

 Diffraction microscopy solves the phase problem using increasingly
sophisticated algorithms based on the support constraint, which
assumes adequate sampling. The object being imaged is limited within a
 support region $S$:
\begin{equation}
\rho(\bm{r})= 0  \text {,   if  $\bm{r}\notin S$}. 
\label{eq:support}
\end{equation}

The sampling conditions required to benefit from the support constraint 
have limited the adoption of  projection algorithms to other
experimental geometries that allow only for sub-Nyquist sampling, 
most notably Bragg sampling from periodic crystalline structures. 

Modern sampling theory however, tells us that that Nyquist sampling
conditions dictated by the support are the worst case scenario for an
arbitrary object. In other words, Shannon was a pessimist: he did not
account for the signal structure.  Compressive sensing theory tells us
that the number of measurements are dictated by the signal structure
rather than it's length.  By structured we mean that the signal has
only a few non-zero coefficients when represented in terms of some
basis, or can be well approximated well by a few non zero
coefficients: they can be described in terms of a few atoms, a few
stars, a few wavelet coefficients, or possibly a few protein folds.
 
In other words, an object of interest is often sparse or
concentrated in a small number of non-zero coefficients in a well
chosen basis, i.e. it can be compressed with no or almost no loss of
information. The meanings of ``well-chosen'' and ``of interest''
are slightly circular: A basis is well-chosen if it succinctly
describes a signal of interest; likewise, a signal is of interest if
it can be described with just a handful of basis
elements. 

Since we do not know where these few terms are located, conventional
wisdom would indicate that one has to first measure the full sample at
the desired resolution, since overlooking an important component of a
signal seems almost inevitable if the whole haystack isn't
thoroughly searched over. There would seem to be no alternative to
processing each signal in its entirety before we can compress it and
store only the desired information (such as the location of the atoms
in a molecule).

But a new theory of ``compressive sampling'' has shown how an image of
interest or structured signals generally can be reconstructed,
exactly, from a surprisingly small set of direct measurements. Cand\`es
and colleagues \cite{candes} 
have defined a notion of ``uniform uncertainty'' that
guarantees, with arbitrarily high probability, an exact solution when
the signal is sparse and a good approximation when it is compressible
or noisy. Their uniform uncertainty condition is satisfied, among
others, by Fourier measurements of a sparse real space object.

  The question of whether modern sensing theory
 is applicable to Fourier amplitude measurements was first 
raised by Moravec, Romberg and
Baraniuk \cite{moravec} who provide an upper bound sampling
condition for the successful retrieval of a sparse signal
autocorrelation, and discuss other conjectures with far reaching 
consequences for low-resolution undersampled phase retrieval.
Since the theory is relatively new and not widely
known to the phase retrieval community, modern sampling theory is
briefly reviewed.

\section{A nonlinear sampling theorem}

The notion that a diffraction pattern from a sparse object can be
reconstructed at sub-Nyquist sampling is not entirely new. The so
called "Direct methods" are routinely used for atomic resolution
imaging of increasing complex molecular structures. Direct methods
enforce the condition that the resulting 
molecule is composed of a finite number of
atoms. The conditions for successful ab-initio phase retrieval using
these methods are strict: it requires (1) atomic resolution and (2)
about 5 strong peaks per atom. Condition (2) means that the algorithms
do not scale well with a large number of atoms since the number of
strong reflections decreases rapidly with the number of atoms. 

Here we look for an alternative answer from modern sampling theory.

Suppose that one collects an incomplete set of frequency samples
(amplitude and phase) of a discrete signal $\rho(r)$ of length $N$. The
goal is to reconstruct the full signal $\rho$ given only $K$ samples
in the Fourier domain where the ``visible frequencies'' are a subset
$\Omega$ (of size $K$) of the set of all frequencies $\{0,\cdots ,
N\}$.

At first glance, solving the underdetermined system of equations
appears hopeless, as it is easy to make up examples for which it
clearly cannot be done. But suppose now that the signal $\rho$ is
compressible, meaning that it essentially depends on a number of
degrees of freedom which is smaller than $N$. Then in fact, accurate and
sometimes exact recovery is possible by solving a simple convex
optimization problem.

\begin{thm} (Candes Romberg and Tao \cite{candes}): 
Assume that $\rho$ is $n_a$ -sparse,
(e.g.  $n_a$ atomic charges in real space with $N$ resolution elements), 
and that we are given $K$ Fourier coefficients with frequencies
selected uniformly at random. Suppose that the number of observations
obeys $K < C n_a  log N$. Then minimizing $\ell_1$ reconstructs $\rho$
exactly with overwhelming probability. In particular, writing $C=22(\delta +
1)$, then the probability of success exceeds $1 - N^{-\delta}$.
\end{thm}

The theorem shows that a simple convex minimization will find the
exact solution without any knowledge about the support, the number of
nonzero coordinates of $r$, their locations, and their amplitudes which
we assume are all completely unknown a priori.

Following \cite{candes} we  formulate this more explicitly. The algorithm that 
optimizes the $\ell_1$ norm:
\begin{eqnarray}
\min |\rho|_1 \text{ subject to }  \FT \rho =F\,,\{ \bm k \in \Omega \}; \\
|\rho|_1=\sum{|\rho(x)|} 
\end{eqnarray}
will find the solution with the correct answer without knowing
 the support $S$ ($S=\{\text{1 if $|\rho_0(x)|>0$, 0
otherwise} \}$), nor the number of non-zero elements $||\rho_0||_0$ 
($||\rho_0||_0=\sum S$). 

In addition, another remarkable result, is that the concept of
``atomicity'' in real space is generalized to sparsity in other basis.
We can use, instead of a dictionary of point atoms, a dictionary of 
curves, beams, or a basis that describes protein folds using a few
terms.
Finally, since the equation above does not depend
strongly on $N$, the number of resolution elements or the basis, 
we can choose redundant basis, with more terms than
real space resolution elements if it helps to describe the object with
fewer terms. 

The only requirement is to be able to find the minimum of
$||\rho||_{1}=\sum |\rho(\bm x)|$. 
These results have already been applied to a number of imaging
techniques, but they require amplitude and phase of the Fourier
coefficients. The question that we try to address here is: what are
the implications to the inversion problem of Fourier magnitude only
recordings?

The following theorem gives an upper  bound:
\begin{thm} (Moravec, Romberg and Baraniuk \cite{moravec}): Assume
that $\rho$ is $n_a$ -sparse,  then it can be recovered exactly 
from a reduced number of random Fourier magnitude measurements 
$K>C n_a^2 \log(4 M/n_a^2)$.
\end{thm}

The theorem is based on the fact that the autocorrelation of an
$n_a$-sparse object is at most $n_a^2$ sparse.  The authors also point
out that $n_a^2$ sparse is worst case scenario. If the object is
connected, then the sparsity of the autocorrelation grows linearly,
instead of quadratically, with the object support to twice the object
size.  In summary, this theorem provides an upper bound for the exact
recovery of the object autocorrelation for an arbitrary but finite
object, as well as from sub-sampled data for an object made of a few
atoms.

The important question then is: does a typical geometry used in x-ray
diffraction satisfy the conditions required to enable the full
recovery of the object's autocorrelation? One of the keys to compressive
sensing is the role played by randomness in the data acquisition
(from the Uniform Uncertainty Principle\cite{candes}). 
Does a Bragg geometry satisfy these conditions?  
In this section we explore the possibility that various sampling geometries
satisfy the condition for compressive recovery of the autocorrelation.

We simulate an object of $n_a$ atoms, and attempt to recover the full
autocorrelation from a subset of Fourier amplitude measurements. The
optimization problem can be stated as  follows:
\begin{equation}
\min |\rho|_1 \text{ subject to }  |\FT a|^2=I_k
\end{equation}
where $a$ is the autocorrelation of the object we are trying to
reconstruct. 

\begin{figure}
\subfigure[Sparse object]
{\includegraphics[width=\figw] {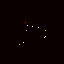} }
\subfigure[Diffr. Patt.]
{\includegraphics[width=\figw]{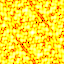} }
\subfigure[Autocorrelation]
{
	\includegraphics[width=\figw]{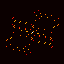}
}
\subfigure[Limited Angle Tomog.]
    {
	\includegraphics[width=\figw]{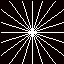}	
    }
\subfigure[Masked FT of (b)]
{
	\includegraphics[width=\figw]{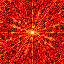}	
}
\subfigure[Recovered autocorr.]
{
	\includegraphics[width=\figw]{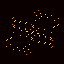}	
}\\
\subfigure[Random Fourier Measurements]
    {
	\includegraphics[width=\figw]{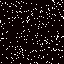}	
    }
\subfigure[Masked FT of (b)]
{
	\includegraphics[width=\figw]{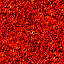}	
}
\subfigure[Recovered autocorr.]
{
	\includegraphics[width=\figw]{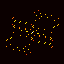}	
}\\
\subfigure[Bragg Sampling]
    {\includegraphics[width=\figw]{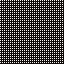}}
\subfigure[Masked FT of (b)]
 {\includegraphics[width=\figw]{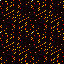}}
\subfigure[Recovered AC]
{
	\includegraphics[width=\figw]{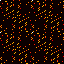}	
}
\subfigure[Random Bragg sampling]
    {
	\includegraphics[width=\figw]{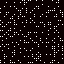}	
    }
\subfigure[Masked FT of (b)]
{
	\includegraphics[width=\figw]{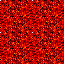}	
}
\subfigure[Recovered AC]
{
	\includegraphics[width=\figw]{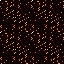}	
}
\subfigure[Radial averaging sampled with oblique lattice]
    {
	\includegraphics[width=\figw]{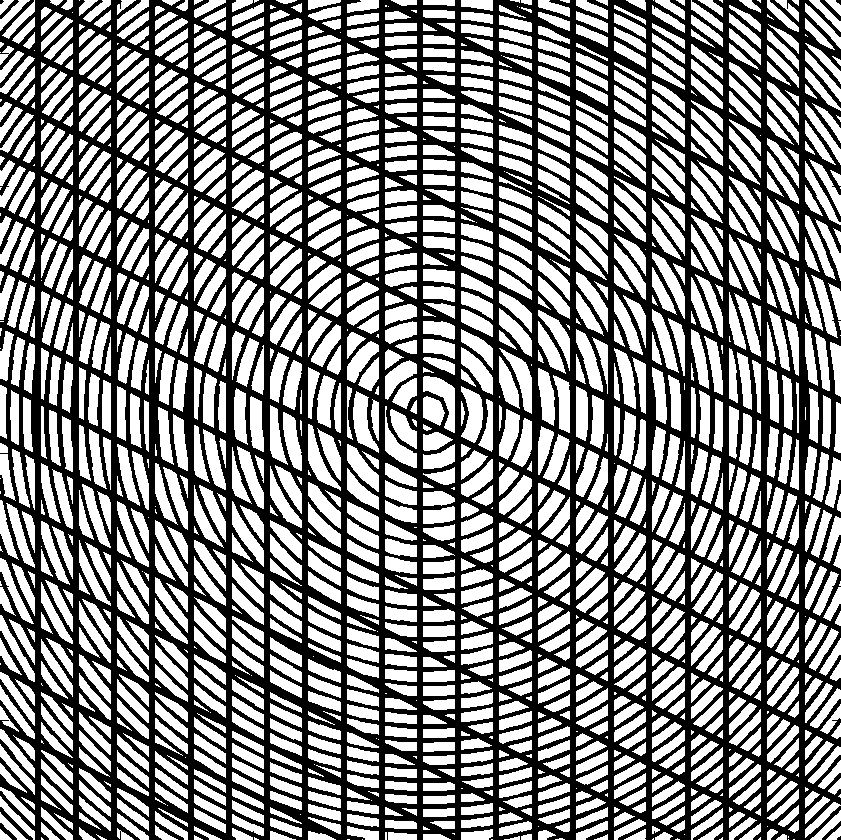}	
    }
\subfigure[FT of radial avg. of (b)]
{
	\includegraphics[width=\figw]{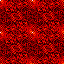}	
}
\subfigure[Recovered AC]
{
	\includegraphics[width=\figw]{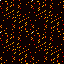}	
}
\caption{
\textbf{Recovering a sparse autocorrelation}.
Top row: (a) Original object, (b) Fourier magnitude (c),
autocorrelation. In the successive rows, Fourier mask (left),
autocorrelation from masked Fourier amplitudes (center), recovered
autocorrelation using $\ell_1$ minimization (right).
Fourier masks: (d) limited angles (g) random (j) Bragg (m) random Bragg (p)
radial average -oblique unit cell.
\label{fig:2}
}
\end{figure}

The problem setting was written within the \textsc{SPARCO}
toolbox\cite{sparco}. With the
addition of radial averaging, the \textsc{SPARCO} toolbox 
provided the operators to simulate a general scattering experiment: 
a Fourier transform, and a Fourier mask. 
The \textsc{SPGL1} software \cite{spgl1} used here was able to
converge quickly to a root.  Various sampling conditions are explored
in Fig. \ref{fig:2}. We can see that random Fourier sampling and limited angle tomography
satisfy the conditions for exact recovery, while Bragg sampling
(sampling every other 2 Fourier components in each dimension) does
not.  However we are able to recover the aliased autocorrelation from
a severely reduced number of measured Bragg reflections or even from
radially averaged powder data (using an oblique unit cell which
causes a reduced number of peak overlaps, in this case 
an average of 10 peak overlap over the same radial shell).

\section{Compressive phase retrieval algorithms}

Although results described above are encouraging, it is obvious that
we can do better than that: we have not utilized the notion that the
function that we have been trying to recover is itself the
autocorrelation of an even sparser object. We therefore look for an algorithm
that recovers a sparse object from subsampled Fourier amplitude
measurements. 

Though problems of this (non-convex) nature are difficult because only
exhaustive algorithms can guarantee convergence, a variety of
heuristics perform well in practice. We follow the notation described
elsewhere \cite{marchesini:rsi} for projection algorithms.

\subsection{Projection algorithms}
The aim of projection algorithms 
is to find a signal that lies in the intersection of two constraint
sets. The first set is that of the measured Fourier magnitudes. 
To compute  the projector corresponding to the Fourier magnitude constraint, 
one first needs to propagate $\rho(\bm r)$ to the data space by a Fourier
transform ${\cal F}$, and then replace estimated amplitudes  $|\trho|$
($\trho(\bm k)={\cal F} \rho(\bm r)$)
with the measured ones, and propagate back to real
space.  Formally, we incorporate the forward ${\cal F}$ and inverse 
${\cal F}^{-1}$ Fourier transform in the operator defined in Fourier
space $\tPm$:
\begin{eqnarray}
\Pm &=&{\cal F}^{-1} \tPm {\cal F}\,,
\end{eqnarray}
and enforce the condition that the image Fourier magnitudes are equal to the
measured ones.
Using these transforms one simplifies the calculation of the projection 
which becomes an element-wise operation on each recovered Fourier
component:
\begin{equation}
\tPm \trho(\bm{k})=\sqrt {I(\bm k )} 
\frac {\trho(\bm k )}{|\trho(\bm k )|},
\end{equation}
This projection requires each Fourier measurement to be sampled individually.
In powder diffraction, partial coherent illumination, broadband
illumination,  the projection
operator needs to be generalized. 
If one defines the averaging operator $A$, the generalization is as
follows \cite{spencepowder}:
\begin{equation}
\tPm \trho(\bm k )=\trho(\bm k )
\sqrt {\frac {I (\bm k_A ) } {A |\trho(\bm k )|^2}},
\end{equation}
where $\bm k_A$ is the corresponding value of the intensity
which $|\trho(\bm k)|$ contributes to. Partial overlap can be 
analyzed using the Richardson-Lucy deconvolution
\cite{richardson,lucy}
 of the intensities under square root (see also 
\cite{fienup:broadband}).
The second projector, used in diffractive imaging, is that of the
support: $\bm{P}_s \rho(\bm{r})=$$\{ \rho(\bm{r})$ if  $\bm{r}\in S$; 
0 otherwise$\}$ that acts element-by-element to the real space
basis. 

While early experiments relied on low resolution
imaging to determine the object support, 
the development of automated support refinement techniques
\cite{shrinkwrap} have enabled diffractive imaging to solve structures
independently. Note that these \textsc{Shrinkwrap} support finding
techniques rely on the most compact (and sparse) object that satisfy
the measurements. 

\subsection{Sparsifying algorithms}
Since we do not know the support of our signal, we replace the
projection onto the support set $P_s$ with a variety of operators $O$ known
to promote sparsity in phase retrieval problems: thresholding,
Sayre' squaring operator \cite{marks},
and a soft thresholding operator that promotes
sparsity through $\ell 1$ optimization \cite{thresholding}
 (Moravec et al. suggest using the
$\ell 1$ norm as a constraint, but this is rarely known):

\begin{eqnarray}
O_{\tau}   x&=& \begin{cases}
\text{$x$ if $|x|>\tau$},\\
 \text {$0$ otherwise.}\\
\end{cases}\\
O_2  x&=& x^2 ,\\
O_{\ell 1} x&=& \begin{cases}
x-\tau \tfrac {x} {|x|}  \text{if $|x|>\tau$,}\\
 \text{$0$ otherwise.}
\end{cases}
\end{eqnarray}
With slight abuse of notation, we use the same symbol for a
projector, and define a reflector operator:
$$
\bm P_{\tau,2,\ell1}=\bm O_{\tau,2,\ell1};
\bm R_{\tau,2,\ell1}=2 \bm O_{\tau,2,\ell1}-\bm I;
$$

 The first most obvious algorithms are simple alternating type:
$$
\rho^{(n+1)}= \Pm \bm O_{(\tau,\ell 1,2)} \rho^{(n)}
$$

We note that the squaring method obtained by alternating the operators
$\Pm O_2 x$ is equivalent  to the ``tangent formula''
of Direct methods, where the squaring operation equation in
real space is performed directly in Fourier space through an
autocorrelation. Remarkable improvements in the range of convergence
has been obtained by increasing the step produced by these algorithms
in real space by 2, using the reflector operator $\bm R$. The charge flipping
algorithm replaces the thresholding operation by with an operator that
moves twice as far:
$$
\rho^{(n+1)}= \Pm \bm R_{(\tau,\ell 1,2)} \rho^{(n)}
$$
 In appendix we describe why it is superior to
simple alternating projections.

\begin{table}
\caption{\label{tab:algorithms}
Summary of various algorithms}
\begin{ruledtabular}
\begin{tabular}{|l|l|}
Algorithm & Iteration $\rho^{(n+1)}=$\\
\hline
\hline
ER & $\bm{P_s P_m}\rho^{(n)}$\\
\hline
SF &$\bm{R_s P_m}\rho^{(n)}$\\
\hline
HIO &\label{eq:HIOtab}
$
\begin{cases}
\bm{P_m} \rho^{(n)}(\bm{r})  & 
	\text {$\bm{r}\in S$} \\
(\bm{I}-\beta \bm{P_m})\rho^{(n)}(\bm r)  & \bm{r}\notin S
\end{cases}$ \\
\hline
DM &
$
\begin{array}{llll}
\{
\bm{I}&+&\beta \bm{P_s} &
 \left [
  \left (
	1+\gamma_s
  \right )
  \bm{P_m}-\gamma_s\bm{I}
 \right ]\\
&-&\beta \bm{P_m} &
 \left [
  \left (
	1+\gamma_m
  \right )
  \bm{P_s}-\gamma_m \bm{I}
 \right ]
 \}
\rho^{(n)} 
\end{array}$\\
\hline
ASR&$\tfrac{1}{2}[\bm{R_s R_m}+\bm{I} ]\rho^{(n)}$\\ 
\hline
HPR&$\tfrac{1}{2}[
	\bm{R_s} 
	\left (
		\bm{R_m}+(\beta-1) \bm{P_m} 
	\right )$\\
&$
	+\bm{I}
	+(1-\beta )\bm{P_m}
	]
\rho^{(n)} 
$\\
\hline
RAAR&$
\left [ \tfrac{1}{2} \beta \left (
	\bm{R_s R_m}+\bm{I} 
	\right )
	+(1-\beta)\bm{P_m}
	\right ] 
\rho^{(n)} $\\
\end{tabular}
\end{ruledtabular}
\end{table}

However we are interested in an algorithm that is more general than
this for reasons that will become apparent in the following
section. In particular we need an algorithm that is robust against 
the relaxation of the positivity of the object, and the change of
basis used to describe the object.

The algorithms tested include: HIO\cite{fienup:josaa82}, 
SF\cite{abrahams:96}, DM \cite{elser:03}, HPR\cite{luke:03} and RAAR\cite{luke:05}
(see \cite{marchesini:rsi} for a review). 
We tested algorithms based on these ideas for
increasingly complex phase retrieval problems. 
A unit cell in a periodic system was filled with a 
limited number of atoms.
First an algorithm has to be stable
around the solution.  If perturbed from the solution, it should go
back or at least not diverge much from it. Perturbations tested
included: distributed noise, ``salt and pepper'' noise and a single
large extra charge added to the structure.  Secondly, it should
converge to the solution starting from an arbitrary set of phases for
a large number of atoms $n_a$. 
 For small $n_a$ all but a handful of algorithms work, as $n_a$ 
increases, it takes longer to converge, and one by one, various 
algorithms stop working. The best
 algorithm is the one that works with highest ratio between number of
atoms and number of measurements. 

Through these numerical tests, the following outperformed all others (fig. 3): 
\begin{eqnarray}
\nonumber
S_1&=&|\Pm \rho-\rho|>\tau_1,\\
\nonumber
S_2&=&\Pm \rho>0\\
\nonumber
 \rho^{(n+1)}&=& \left (S_1 \& S_2 \right )  \left (\Pm \rho^{(n)}
\right ) + (1-S_1) \left (\rho^{(n)}\right ) \\ &-&\beta \Pm \rho^{(n)};
\label{eq:espresso}
\end{eqnarray}
where  $>$ is intended as a relational operator, $S_{1,2}$ are binary
masks.  If positivity cannot be
enforced, then $S_2=1$.
From here on, this algorithm will be referred to as \textsc{Espresso}, in honor of the compressed coffee.

\begin{figure}
	\includegraphics[width=.4\textwidth]{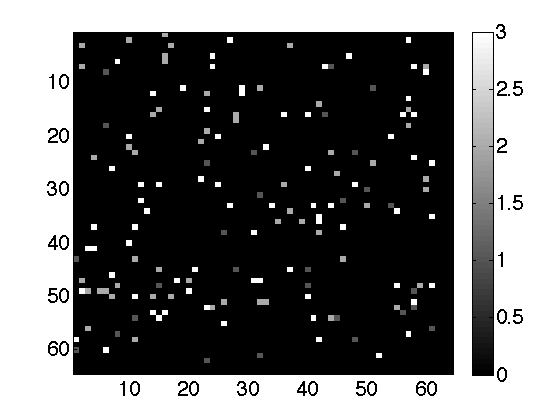}	
	\includegraphics[width=.4\textwidth]{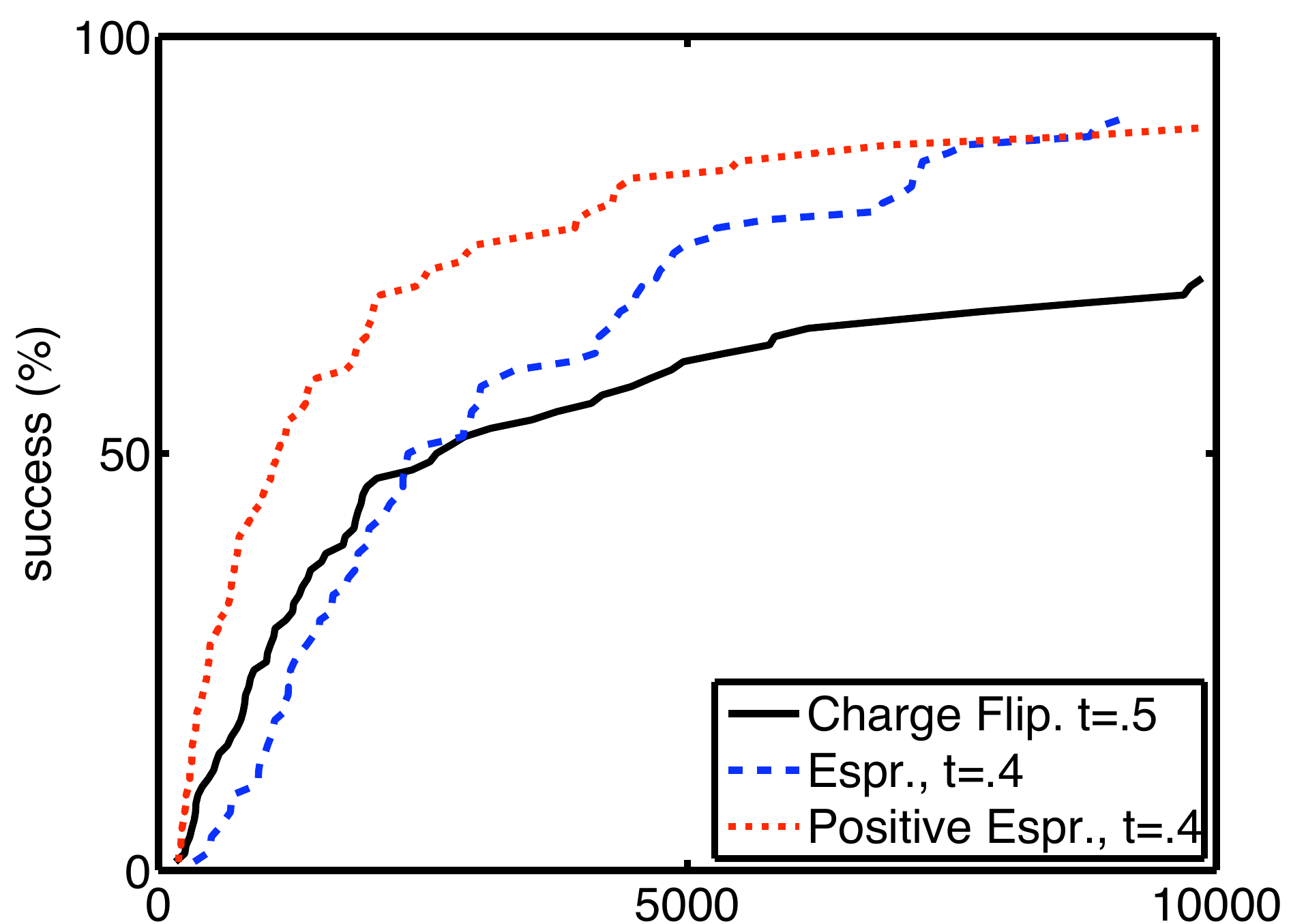}	

\caption{
\textbf{Recovering a sparse object in real space}.
Comparison of various algorithms. Top: test unit cell, 64$\times$64
resolution elements, 160 atoms of values ranging from 1 to 3
(arbitrary units). Bottom: success rate vs number of iterations, using
the optimal threshold level ($\tau=.5$ for charge flipping, $.4$ for 
\textsc{Espresso}); and $\beta=.5$. 
\label{fig:3}
}
\end{figure}

\section{Beyond atoms}
The algorithms described perform remarkably well in case an object
being imaged is sparse in real space, namely that it is composed of a
limited number of atoms.  Object made of more than a few atoms, such as
large macromolecules or biological cells, require a different approach.

Natural objects are often characterized by the fact that they are
compressible in some basis. Everyone is familiar with the fact that 
images of millions of pixels can be saved with nearly indistinguishable
accuracy at a small fraction of the initial image size. 
In other words, natural objects can often be accurately described 
in terms of only a few non-zero coefficients in some basis. 

 Moravec et al.\cite{moravec} conjecture that compressive methods for
phase retrieval could be applicable in a basis other than the real
space. Here we set out to test this conjecture for the
crystallographic phase problems with the \textsc{Espresso} algorithm.
Will all basis work? not if they are too localized in
Fourier space, as preliminary tests with curvelets \cite{curvelets} 
seem to suggest. 

The problem in this test is to recover a Schepp-Logan phantom test
object. As in many tomographic settings, we seek to sparsify the
 the gradient of the object.  

The discrete gradient is obtained by simple matrix operations.  One defines a
two-dimensional matrix $D_x$ of the same size as $\rho$, with the
first two elements equal to -1 and +1 respectively, and the rest of
the elements equal to 0. The discrete derivative is obtained by a
simple convolution between the object and this matrix: $\partial_x
\rho(\bm r)=D_x *\rho(\bm r)$.  Convolution becomes a product in
Fourier space, where we define the discrete Fourier transform 
$\tilde D_x={\cal F}D_x$: $\partial_x \rho={\cal F}^{-1} (\tilde D_x \tilde
\rho)$. Similar arguments apply for the other direction. We define the
matrix $D_y=D_x^{T}$ as the transpose of $D_x$. 

The inverse operation used to recover the object from the compressed
gradient values ($\partial_x \rho, \partial_y \rho$) was  the
following peudoinverse:
\begin{equation}
\rho={\cal F}^{-1}
\frac{\tilde D_x^\dagger {\cal F} \partial_x \rho+\tilde D_y^\dagger
{\cal F} \partial_y \rho}
{\tilde D_x^\dagger \tilde D_x+\tilde D_y^\dagger \tilde D_y+\eps,}
\end{equation}
with $\eps$ a regularization term.
Once we have obtained
the gradient, $[\partial_x, \partial_y] \rho$, we apply the Espresso
algorithm in this space, trying to compress the gradient.
The norm used was $\sqrt{|\partial_x \rho|^2+|\partial_y \rho|^2}$:
\begin{eqnarray}
\nonumber
S_1&=&\sqrt{|\partial_x \Pm \rho-\rho|^2+|\partial_y \Pm \rho-\rho|^2}>\tau_1,\\
\nonumber
S_2&=&\sqrt{|\partial_x \Pm \rho|^2+|\partial_y \Pm \rho-\rho|^2}>\tau_2,\\
\nonumber
\partial_x \rho^{n+1}&=& (S_1 \& S_2)  (\partial_x \Pm \rho^{n}) + (1-S_1)
(\partial_x \rho^{n}) -\beta \partial_x \Pm \rho^{n},\\
\nonumber
\partial_y \rho^{n+1}&=& (S_1 \& S_2)  (\partial_y \Pm \rho^{n}) + (1-S_1) (\partial_y \rho^{n}) -\beta \partial_y \Pm \rho^{n};
\label{eq:espresso}
\end{eqnarray}
although the two gradient components could be treated separately, 
each with their own supports $S_{1,2}$.

Examples are shown in Fig. 4, where different masks are applied to the
Fourier data: Full sampling, missing central region due to the
beamstop, limited angles of tomographic datasets, Bragg sampling, and
radially averaged powder data. Notably, the periodic crystal symmetry
is ``discovered'' by sampling the diffraction pattern at the Bragg condition.
In the case of powder diffraction, with an average of 20 overlaps per
sampled point, the algorithm is still able to recover the outline of
the object, or the dominant terms contributing to the gradient. The
unit cell used for calculating the radial average was oblique, to
reduce some peak overlaps. In each case the threshold is defined
dynamically:
$\tau=2\sigma$, with $\sigma=rms(|\rho-\Pm \rho|)$.

\begin{figure}
\subfigure[Fourier mask, full sampling]
{\fbox{\includegraphics[width=\figwIV]{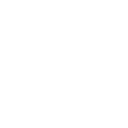}}}
\subfigure[Recovered object]
{\includegraphics[width=\figwIV]{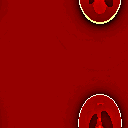}}
\subfigure[F. Mask, beamstop]
{\fbox{\includegraphics[width=\figwIV]{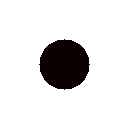}}}
\subfigure[Recovered object]
{ \includegraphics[width=\figwIV]{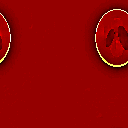}	 }
\subfigure[F. Mask, limited angle tomog.]
{\includegraphics[width=\figwIV]{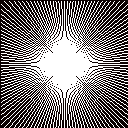}}
\subfigure[Recovered object]
{\includegraphics[width=\figwIV]{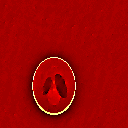}	}
\subfigure[F. Mask: Bragg sampling]
{\includegraphics[width=\figwIV]{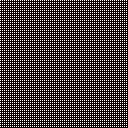}}
\subfigure[Recovered object]
{\includegraphics[width=\figwIV]{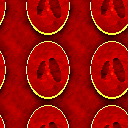}}
\subfigure[Powder pattern-log scale]
{\includegraphics[width=\figwIV]{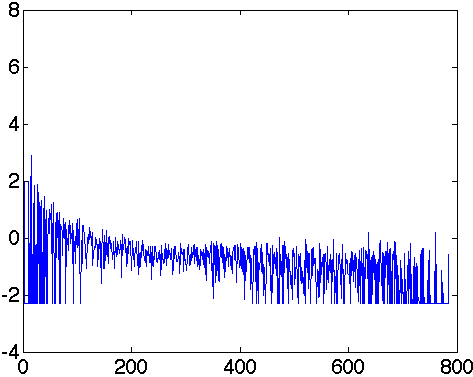}}
\subfigure[Recovered object]
{\includegraphics[width=\figwIV]{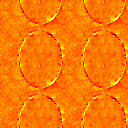}}
\caption{
\textbf{Recovering an object without atomic resolution}.
Shepp-Logan phantom is embedded into a 128$\times$128 image.
Left: Fourier mask, Right: reconstruction. 
(a)-(b): Full sampling, (c)-(d) Beam stop, (e)-(f) Missing angles,
(g)-(h) Bragg sampling, (i)-(j) Powder sampling, 20 peak overlaps on
average.
\label{fig:4}
}
\end{figure}

\section{Conclusions}
The revolutionary findings of compressive sensing
have shown how severely incomplete data can be used to recover 
both sparse and compressible signals from Fourier amplitude and phase
measurements. This paper discusses its implications for the phase
retrieval problem, namely that one does not need to assume a finite
signal bandwidth, nor atomic resolution. 
A new algorithm for compressive phase retrieval has been proposed. It
outperforms existing algorithms for atomic resolution data, and 
can be applied at arbitrary resolution to crystal diffraction,
provided that the sample is compressible in some basis. 
Optimization of the parameter $\beta$ on the fly at each iteration
\cite{marchesini:josaa07}, promises to further improve this
algorithm and  will be subject of future work. 
The search for the optimal sparsifying basis and improved compressive
algorithms  promises to revolutionize low
resolution phasing in X-ray diffraction.

\textsc{Acknowledgments} 
This work was performed under the auspices of the U.S. Department of
Energy by the Lawrence Berkeley National Laboratory under Contract
No. W-7405-EEG-48 and the Director, Office of Energy Research.
Part of this work was done while visiting the Center for Free Electron
Lasers at DESY. Motivation for this work is due to the serial
crystallography project of J. C. H. Spence et al., and recent
experimental results by D. Shapiro et. al., as well as
the diffractive imaging program   at the Advanced Light Source. 
Inspiration for this work is due to discussions with C. Yang at the 
computational research division of Lawrence Berkeley Lab.

\appendix
\section{Charge flipping}
Take a known object with charge density distribution $\rho(\bm r)$, with
Fourier amplitude $|\trho(\bm k)|=\sqrt{I(\bm k)}$. 
We first look at perturbations
from the solution. If we add a random noise distribution $a$ with root
mean square (r.m.s.) $\sigma< \min(\rho: \rho>0) $, a simple threshold will be
able to isolate the signal $\rho>\sigma$ from the noise.  As the noise
grows, an increasing large percentage of noise will survive the
thresholding operation and the thresholding algorithm will stagnate at
first local minimum it encounters.  At low perturbation levels, a simple
thresholding operation is the obvious and effective answer. Flipping
components below a threshold, rather than setting them to zero would
just flip sign within a first order term, and at best remove second 
order $O(a^2/||\rho||)$ contributions to the noise $a$.  
We consider  ``salt and pepper'' noise
  $a=\sum_{i=1}^s a_i^s \delta(x-x_i)$ 
with a few $s$ terms, such as if we displaced a small percentage of the peaks 
from the correct answer). For simplicity, let us consider 
just a single charge $a$ in $x_0$:
\begin{equation}
\rho'=\rho+a \delta (x-x_0);
\end{equation}
We enforce the measured data  by applying the Fourier magnitude
projector to the current guess 
$\trho'={\cal{F}} \rho'= \trho+a e^{ik x_0}$:

The projection is performed by Fourier transforming $\trho'={\cal F} \rho'$, 
and enforcing the Fourier magnitude:
\begin{eqnarray}
\nonumber
\tPm \trho'&=&\sqrt{\frac {I}{|\trho'|^2}} \trho'\\
\nonumber
&=&\sqrt{\frac {I}{|\trho|^2}} 
\frac {\trho+a e^{ik x_0}} 
{
\sqrt{ 
	1+
	\tfrac{\trho^*}{|\trho|^2} a e^{ik x_0}+
	\tfrac{\trho}{|\trho|^2} a^* e^{-ik x_0}+
	\tfrac{|a|^2}{|\trho|^2}
	}
}
\end{eqnarray}
using Taylor series $\tfrac 1 {\sqrt{1+x}}=1-\half  x +\tfrac{3}{8} x^2$:
\begin{eqnarray}
\nonumber
&=&\left ( \trho+a e^{ik x_0} \right )
\left ( 
	1-
	\half \tfrac{\trho^*} {|\trho|^2} a e^{ik x_0}-
	\half \tfrac{\trho}{|\trho|^2} a^* e^{-ik x_0}+O(a^2)
\right )\\
\nonumber
&=& 
 \trho+\half a e^{ik x_0} 
	-\half \tfrac{\trho \trho}{|\trho|^2} a e^{-ik x_0}+O(a^2)
\\
\nonumber
	 &=& \trho+\half a e^{ik x_0} -\half a^* F(x_0) + O(a^2)
\end{eqnarray}
with $F(x_0)=  \tfrac{\trho }{\trho^*}  e^{-ik x_0}$.
The term $\frac{\trho}{\trho^*}$ has a rapidly oscillating phase, and diffuses
half of the charge $a$  everywhere in the unit cell through the function 
$ f= {\cal F}^{-1} F $.
At first order in $a$ one
obtains:
\begin{equation}
\Pm \rho'= \rho+\half  a \delta (x-x_0) -\half a^* f(x_0) +O(a^2)
\label{eq:2}
\end{equation}
with $f(x_0)$ being distributed over the unit cell.
Since the charge is reduced only by half, we could try moving twice as
far, instead of moving from $rho$ to $\Pm rho$ by a step $I- \Pm$, using a
reflector  $\bm{R}=2\bm{P}-\bm{I}$ to:
\begin{equation}
\bm{R_m} \rho'= \rho - a^* f(x_0) +O(a^2)
\end{equation}

Suppose the threshold  only picks up the term $\half a^*f(x_0)$, then
the flipping operation converges to the solution in one step (at first
order):
\begin{eqnarray}
\Pm [ \rho+\half a \delta (x-x_0) +\half a^*f(x_0)]= \rho+O(a^2)
\end{eqnarray}
Although this result describes only local convergence, it shows how
a few wrong peaks can be recovered easily.


\begin{thebibliography}{27}
\bibitem{homometric}
L. Pauling and M. D. Shappell, Zeits. f. Krist. \textbf{75}, 128 (1930).
\bibitem{Perutz:1938} J.D. Bernal, I. Fankuchen, M. F. Perutz,  ``An
X-Ray Study of Chymotrypsin and Haemoglobin.'' Nature 141, 523-524
(1938). 
\bibitem{Sayre:1952} D. Sayre, ``On the implication of a theorem due
to Shannon'', Acta Cryst. \textbf{5}, (1952) 843.
\bibitem{bruck} Y. M. Bruck and L. G. Sodin. 
``On the ambiguity of the image reconstruction problem.'' 
Optics Communications, \textbf{30}(3):304-308, (1979). 
\bibitem{bates} R. Bates. ``Fourier phase problems are 
uniquely solvable in more than one dimension. I: Underlying theory'' 
Optik, \textbf{61}(3):247-262, 1982. 
\bibitem{hayes1}
M. H. Hayes, ``The reconstruction of a multidimensional sequence from
the phase or magnitude of its Fourier transform,'' IEEE Trans. 
ASSP \textbf{30}(2), 140-154, (1982).
\bibitem{hayes2} M. H. Hayes and J. H. McClellan, ``Reducible polynomials
in more than one variable,'' Proc. IEEE \textbf{70}(2),  197-198, (1982).
\bibitem{miao:nature} 
J. Miao, P. Charalambous, J. Kirz, D. Sayre, ``Extending the
methodology of X-ray crystallography to allow imaging of
micrometre-sized non-crystalline specimens,'' Nature \textbf{400},
342-344 (1999).
\bibitem{shapiro:PNAS05} D. Shapiro, P. Thibault, T. Beetz, V. Elser,
M. Howells, C. Jacobsen, J. Kirz, E. Lima, H. Miao, A. Neiman,
D. Sayre, 
``Biological imaging by soft x-ray diffraction microscopy,''
Proc. Nat. Acc. Sci.  \textbf{102}, 1543-1546 (2005).
Nature \textbf{442}, 63-67 (2006).
\bibitem{barty:PRL08} 
A. Barty,
S. Marchesini, H. N. Chapman, C. Cui, M. R. Howells, D. A. Shapiro,
A. M. Minor, J. C. H. Spence, U. Weierstall, J. Ilavsky, A. Noy,
S. P. Hau-Riege, A. B. Artyukhin, T. Baumann, T. Willey, J. Stolken,
T. van Buuren, J. H. Kinney, 
``Three-dimensional coherent X-ray diffraction imaging of a ceramic nanofoam: 
determination of structural deformation mechanisms,'' 
\prl \textbf{101}, 055501 (2008), [arxiv:0708.4035].
\bibitem{spence:book} 
 P.~W.~Hawkes \&  J.~C.~H.~Spence (Eds.), 
\textit{Science of Microscopy} (Springer, 2007).
\bibitem{candes}
E. J. Cand\`es, J. Romberg and T. Tao, ``Robust uncertainty principles:
exact signal reconstruction from highly incomplete frequency
information,'' IEEE Trans. Inform. Theory, \textbf{52},
489-509 (2006) [arXiv:math/0409186].
\bibitem{moravec} 
M. L. Moravec, J. K. Romberg, R. G.  Baraniuk, Richard,
Compressive phase retrieval, 
Wavelets XII. Proc. SPIE \textbf{6701}, 670120 (2007).
\bibitem{sparco}
E. van den Berg and M. P. Friedlander,
SPGL1: A solver for large-scale sparse  reconstruction,
http://www.cs.ubc.ca/labs/scl/index.php/Main/Spgl1
\bibitem{spgl1}
E. van den Berg and M. P. Friedlander, "Probing the Pareto frontier
for basis pursuit solutions", UBC Computer Science Technical Report
TR-2008-01, January 2008.  Available at
http://www.optimization-online.org/DB\_F
\bibitem{richardson} W. H. Richardson, ``Bayesian-Based Iterative
Method of Image Restoration". \josaa \textbf{62}, 55-59 (1972). 
\bibitem{lucy} L. B. Lucy, ``An iterative technique for the
rectification of observed distributions". Astronomical Journal 
textbf{79}  745-754 (1974).
\bibitem{fienup:broadband} J.R. Fienup, 
``Phase Retrieval for Undersampled Broadband Images,'', 
\josaa \textbf{16}, 1831-1839 (1999).
\bibitem{shrinkwrap} 
S. Marchesini, H. He, H. N. Chapman, S. P. Hau-Riege,
A. Noy, M. R. Howells, U. Weierstall, J.C.H. Spence,
``X-ray image reconstruction from a diffraction pattern alone,''
\prb \textbf{68},  140101(R) 1-4, (2003), [arXiv:physics/0306174].
\bibitem{marks} 
L. D. Marks, W. Sinkler and E. Landree, ``A Feasible Set Approach to
the Crystallographic Phase Problem'', 
Acta Cryst. \textbf{A55}, 601-612 (1999).
\bibitem{Fienup:1978} J. R. Fienup, 
``Reconstruction of an Object from the Modulus of Its Fourier Transform,''
\ol  \textbf{3},  27-29  (1978).
\bibitem{fienup:josaa82} J. R. Fienup, ``Phase retrieval algorithms: a
comparison'', \ao  \textbf{21},  2758-2769 (1982).
\bibitem{abrahams:96}  J.~P.~Abrahams, A.~W.~G.~Leslie, Acta Cryst. 
\textbf{52}, 30-42 (1996).
\bibitem{elser:03} V.~Elser, 
``Phase retrieval by iterated projections,'' \josaa \textbf{20}, 40-55 (2003). 
\bibitem{luke:03}  H.~H.~Bauschke, P. L. Combettes, and D. R. Luke,
``Hybrid projection reflection method for phase retrieval,'' 
\josaa  \textbf{20}, 1025-1034 (2003).
\bibitem{luke:05} D.~R.~Luke,
``Relaxed Averaged Alternating Reflections for Diffraction Imaging,''
 Inverse Problems  \textbf{21}, 37-50
(2005)., (arXiv:math.OC/0405208).
\bibitem{combettes} P. L. Combettes, 
``The Convex Feasibility Problem in Image Recovery, in Advances in
Imaging and Electron Physics,'' (P. Hawkes, Ed.), vol. 95,
pp. 155-270. (Academic Press,New York 1996).
\bibitem{marchesini:rsi} S. Marchesini, 
``A unified evaluation of iterative projection algorithms for 
phase retrieval,'' Rev. Sci. Inst. \textbf{78}, 011301 1-10 (2007),
[arXiv:physics/0603201].
\bibitem{oszlanyi} G. Oszl\'anyi and A. S\"uto, 
``Ab initio structure solution by charge flipping,''
Acta Cryst.  \textbf{A60}, 134-141 (2004) [arXiv:cond-mat/0308129].
\bibitem{spencepowder} J. Wu, K. Leinenweber, J.~C.~H. Spence, 
``Ab initio phasing of X-ray powder diffraction 
patterns by charge flipping,'' 
Nature Materials \textbf{5}, 647-652 (2006).
\bibitem{curvelets}
E. J. Cand\'es, 
D. L. Donoho,  ``New tight frames of curvelets and optimal 
Representations of objects with piecewise C$^2$ singularities.'' Comm. Pure
Appl. Math. \textbf{57}, 219-266 (2004).
\bibitem{marchesini:josaa07} S. Marchesini, \josaa 24,32890-3296
(2007), [arXiv:physics/0611233].
\bibitem{}
G. Oszl\`anyi and A. S\"uto, ``Ab initio structure solution by charge
flipping. II. Use of weak reflections,'' Acta Crystallogr. \textbf{A61}, 147-152
(2005).
\bibitem{thresholding} I. Daubechies, M. Defrise, and 
C. D. Mol, ``An iterative thresholding algorithm for linear inverse problems 
with a sparsity constraint,'' Comm. Pure Appl. Math. \textbf{57}(11),
1413-1457, (2004).
\end{thebibliography}
\end{document}